\documentclass[12pt]{article}

\title{Phantom Energy Accretion onto Black Holes in Cyclic Universe}
\author{\textbf{Cheng-Yi Sun\footnote{cysun@mails.gucas.ac.cn; ddscy@163.com}\ $^{,a}$\
}\\ \\
 {$^a$\small Institute of Modern Physics, Northwest University,}\\
     \small Xian 710069, P.R. China.}

\begin{document}
\maketitle
\begin{abstract}
Black holes pose a serious problem in the cyclic or oscillating
cosmology. It is speculated that, in the cyclic universe with
phantom turnarounds, black holes will be torn apart by the phantom
energy before turnaround before they can create any problems. In
this paper, using the mechanism of the phantom accretion onto black
holes, we find that black holes do not disappear before the phantom
turnaround. But the remanent black holes will not cause any problems
due to the Hawking evaporation.

\end{abstract}
\section{Introduction}
The scenario of cyclic or oscillating universes is an attractive
idea in the theoretical cosmology since it is expected to avoid the
initial sigularity by providing an infinitely oscillating universe.
This idea has a long history \cite{Friedmann,Tolman}. In recent
years, there have been many discussions on such a topic
\cite{Steinhardt_Turok,a0405353,t0610213}. Generally, however,
cyclic models of the universe confront a serious problem: black
holes.
If black holes formed during expansion of the universe survive into
the next cycle, they will grow even larger from cycle to cycle and
act as defects in an otherwise nearly uniform universe. Eventually,
the black holes will occupy the entire horizon volume, and then the
cyclic models break away.

In \cite{a0405353}, by assuming the existence of the phantom dark
energy and using the modified Friedmann equation, the authors
suggested an oscillating cosmology. It is argued that the any black
holes produced in an expanding phase in the universe are torn apart
before they can create problems during contraction. A rough
calculation has been given. In general relativity, the source for a
gravitational potential is the volume integral of $\rho+p$, where
$\rho$ is the energy density in the universe and $p$ is the
pressure. So an object of radius $R$ and mass $M$ is pulled apart
when $-(4\pi/3)(\rho+p)R^3\sim M$. Then a black hole of mass $M$ and
horizon radius $R=2GM$ is pulled apart when the energy density of
the universe has climbed up to a value
$\rho_{BH}\sim(3/32\pi)(M^2G^3|1+w|)^{-1}$, where $p=w\rho$ is the
equation of state ( for phantom dark energy $w<-1$). Then black
holes will be torn apart before turnaround, if $\rho_{BH}<\rho_c$,
where $\rho_c$ is the critical energy density in the cyclic model,
namely the energy density corresponding to the turnaround (and
bounce).

However, it is obvious that the destruction of black holes is not an
instantaneous even just happened at $\rho\sim\rho_{BH}$, but a
process. At the same time, the qualitative analysis in
\cite{a0405353} is too rough and can not be taken as a mechanism of
tearing up black holes. We needs such a mechanism in order to know
whether the analysis in \cite{a0405353} does work or not. In
\cite{g0402089}, the authors have suggested a mechanism in which, by
accreting the phantom energy, the mass of a black hole decreases at
the rate $\dot{M}=4\pi AM^2(\rho+p)$, where $A$ is a positive
dimensionless constant. Replacing $\rho$ and $p$ by the effective
energy density $\rho_{eff}=\rho(1-\rho/\rho_c)$ and the effective
pressure $p_{eff}=p(1-2\rho/\rho_c)-\rho^2/\rho_c$ respectively, the
author of \cite{0708.1408} used this mechanism, $\dot{M}=4\pi
AM^2(\rho_{eff}+p_{eff})$, to study the destruction of black holes
in cyclic models. The conclusion in \cite{0708.1408} is that, in the
expanding stage of the universe, through the phantom accretion, the
masses of black holes first decrease and then increase. And at the
turnaround, black hole masses restore their initial values. So it is
claimed that black hole cannot be torn up in the cyclic model of
\cite{a0405353}.

But in \cite{0709.1630}, the author argued that $\rho_{eff}$ and
$p_{eff}$ are not proper physical quantities. Taking into account
this view, in this paper, we will study the destruction of black
holes in the phantom cyclic universe by using $\dot{M}=4\pi
AM^2(\rho+p)$. Similar application in the brane cosmology has been
discussed in \cite{0709.4410}.

The paper is organized as follows. In section \ref{Accretion}, we
analyze the phantom energy accretion onto black holes in the cyclic
model of \cite{a0405353}. Section \ref{Conclusion} contains
conclusion and discussion.

\section{Accretion of Phantom Fluid in Cyclic Models}
\label{Accretion} Here, we consider that the dark energy fluid,
covers the whole space in the homogeneous and isotropic form with
the dark energy density $\rho$ and pressure $p$. For an asymptotic
observer, the black hole mass $M$ changes at the rate
\cite{g0402089}
\begin{equation}
  \label{dMdt}
  \dot{M}=4\pi AM^2(\rho+p).
\end{equation}
Here and after, the overhead dot denotes the derivative with respect
to the cosmic time and $G=c=1$. Assuming the universe dominated by
the dark energy, the expansion of the universe is governed by the
Friedmann equation
\begin{equation}
  \label{Friedmann}
  H^2=\frac{8\pi}{3}\rho
\end{equation}
and the local energy conservation law of the dark energy
\begin{equation}
  \label{conservationLaw}\dot{\rho}+3H(\rho+p)=0,
\end{equation}
where $H\equiv\dot{a}/a$ is the Hubble parameter. For the phantom
dark energy with the equation of state $w=p/\rho<-1$, $\rho\propto
a^{-3(1+w)}$ increases as the expansion of the universe. In order to
avoid the big rip \cite{Caldwell}, we use the modified Friedmann
equation
\begin{equation}
  \label{ModFriedmann}
  H^2=\frac{8\pi}{3}\rho(1-\frac{\rho}{\rho_c}),
\end{equation}
where $\rho_c$ is the critical energy density about at the order of
the Planck density. The modified Friedmann equation
(\ref{ModFriedmann}) has been suggested from different set-ups
\cite{brane,QG,LQG}. Then, in the expanding phase of the universe,
the phantom energy density $\rho$ increases. When $\rho=\rho_c$, due
to Eq.(\ref{ModFriedmann}), a turnaround occurs. After the
turnaround, the universe begins to contract. In the contracting
phase, the energy densities of other non-phantom components in the
universe increase, and eventually dominate the evolution of the
universe. When the dominant energy density reaches the critical
energy density $\rho_c$ again, a bounce occurs. Roughly, this is the
scenario of an oscillating cosmology in \cite{a0405353}. Now let us
study the evolution of black hole masses as the phantom energy
accretion in the scenario.

\subsection{Before Turnaround}
Using the two equations (\ref{dMdt}) and (\ref{conservationLaw}), we
get
\begin{equation}
  \label{dMdRhoH}
  \frac{dM}{M^2}=-\frac{4\pi A}{3H}d\rho.
\end{equation}
Before the turnaround, we have
\begin{equation}
  \label{+H}H=\sqrt{\frac{8\pi}{3}}\sqrt{\rho(1-\frac{\rho}{\rho_c})}.
\end{equation}
Substituting this equation into Eq.(\ref{dMdRhoH}), we get
\begin{equation}
  \label{dMdRho}
  \frac{dM}{M^2}=-\frac{D}{\sqrt{\rho(1-\rho/\rho_c)}}d\rho,
\end{equation}
with $D\equiv\sqrt{\frac{2\pi}{3}}A$. The integration of
(\ref{dMdRho}) gives
\begin{equation}
  \label{MbeforT}
  M=\frac{M_i}{1+2DM_i\sqrt{\rho_c}(\arcsin{\sqrt{\rho/\rho_c}}
  -\arcsin{\sqrt{\rho_i/\rho_c}})},
\end{equation}
with $\rho_i\leq\rho\leq\rho_c$. Here $M_i$ and $\rho_i$ denote
respectively the black hole mass and the phantom energy density at
the moment when the phantom energy begins to dominate the evolution
of the universe. Generally, $\rho_i\ll\rho_c$. Then we obtain
\begin{equation}
  \label{Mrho}M\simeq\frac{M_i}{1+2M_iD\sqrt{\rho_c}\arcsin{\sqrt{\rho/\rho_c}}}.
\end{equation}
So the black hole mass at the turnaround is
\begin{equation}
  \label{Mc}
  M_c\simeq\frac{M_i}{1+\pi M_iD\sqrt{\rho_c}}
\end{equation}
This result means that black holes, by accreting the phantom energy,
do not disappear before the turnaround. But our result is different
from the result of \cite{0708.1408}. In \cite{0708.1408}, it is
claimed that, through the phantom accretion, black hole mass will
decrease first, and then increase until restoring its initial mass
at the turnaround. Here, our result, Eq.(\ref{MbeforT}), indicates,
through the phantom accretion, the black hole mass always decreases
as the expansion of the universe, and, at turnaround, reaches the
minimum $M_c$ in the expanding phase. For $M_i\gg M_p= G^{-1/2}$,
$M_c$ becomes independent of $M_i$
\begin{equation}
  \label{Mcs}
  M_c\simeq\frac{1}{\pi D\sqrt{\rho_c}}
\end{equation}

\subsection{After Turnaround}
After turnaround, the universe begins to contract and the phantom
energy density $\rho$ decreases as the contraction.
Eq.(\ref{MbeforT}) cannot be used directly because $H$ is negative
in the contracting phase of the universe. Substituting the equation
\begin{equation}
  \label{-H}
  H=-\sqrt{\frac{8\pi}{3}}\sqrt{\rho(1-\frac{\rho}{\rho_c})}
\end{equation}
into Eq.(\ref{dMdRhoH}), we obtain
\begin{equation}
  \label{-dMdRho}
  \frac{dM}{M^2}=-\frac{D}{\sqrt{\rho(1-\rho/\rho_c)}}d\rho.
\end{equation}
The integration of the equation gives
\begin{equation}
  \label{MafterT}
  M=\frac{M_c}{1+DM_c\sqrt{\rho_c}(\pi-2\arcsin{\sqrt{\rho/\rho_c}})},
\end{equation}
with $\rho\leq\rho_c$. Then Eq.(\ref{MafterT}) shows that, after
turnaround, as the universe contracting, the phantom energy density
decreases and the black hole masses continue to decrease. When
$\rho\ll\rho_c$, the black hole mass is
\begin{equation}
  \label{Mf}M_f\simeq\frac{M_c}{1+\pi DM_c\sqrt{\rho_c}}
\end{equation}
For $M_i\gg M_p$, using Eq.(\ref{Mcs}), we find the final mass of
black holes is
\begin{equation}
  \label{Mfs}M_f\simeq\frac{M_c}{2}\simeq\frac{1}{2\pi D\sqrt{\rho_c}}
\end{equation}

\subsection{Destruction of Black holes}
The analysis above shows that, by accreting the phantom energy, a
black hole in the cyclic universe with phantom turnaround does not
disappear, but has a remanent mass $M_c$ at the turnaround. This
means, through the phantom energy accretion, black holes in the
cyclic model of \cite{a0405353} can not be pulled apart before
turnaround. Then it seems that the argument of destruction of black
holes in \cite{a0405353} is wrong and the problem of black holes
still stand.

However, it has been argued in \cite{a0405353} that for a black hole
with mass $M=10^5M_p$, it Hawking evaporates in a time
$\tau\sim(25\pi M^3/M_p^4)\sim10^{-27}$sec and causes no problems.
Let us estimate the value of $M_c$. We have taken $G=M_p^{-2}=1$ in
the derivations above. Keeping $M_p$ explicitly, we can rewrite
Eq.(\ref{Mc}) as
\begin{equation}
  \label{McMp}M_c\simeq\frac{M_i}{1+\pi M_iD\sqrt{\rho_c}/M_p^3}.
\end{equation}
$D$ may be taken as a constant of order unity and $\rho_c\sim
{M_p}^4$. Then we have $M_c\sim\frac{M_i}{1+M_i/M_p}$.  This
implies, for a black hole with $M_i\gg M_p$, the remanent mass $M_c$
is about at the order of the Planck mass $M_p$ and the remanent
black hole Hawking evaporates in a time $\tau\sim10^{-43}$sec at the
order of the Planck time. Then, fortunately, the remanent black
holes do not cause problems. Now, we know, in the cyclic model with
phantom energy turnarounds, black hole masses decrease due to the
phantom energy accretion. Before a turnaround, black holes can not
be torn apart, but the remanent black holes with masses $M_c\sim
M_p$ are left at the turnaround. However, the remanent do not cause
problems in the cyclic model because of the Hawking evaporation.

\section{Conclusion and Discussion}
\label{Conclusion}

In the cyclic or oscillating cosmology, black holes pose a serious
problem. In \cite{a0405353}, an oscillating cosmology with phantom
energy turnarounds is suggested and it is argued roughly that black
holes are torn apart before the turnaround. In \cite{g0402089}, a
successful mechanism in which the black hole masses decrease due to
the phantom energy accretion is obtained. In this paper, by using
the result of \cite{g0402089}, we have surveyed the destruction of
black holes in the cyclic cosmology with phantom energy turnarounds.

Similar work has been done in \cite{0708.1408}. Their conclusion is
that, through the phantom accretion, black hole mass will decrease
first, and then increase until restoring its initial mass at the
turnaround. Then the author claimed the problem of black holes still
stands in the cyclic model with phantom turnaround. However, this
conclusion is obtained by using the effective energy density and
pressure which are unphysical variables \cite{0709.1630} , rather
than the energy density and pressure. Using the energy density and
pressure, we find that, due to the phantom energy accretion, the
mass of a black hole always decreases before turnaround, and at
turnaround reaches the remanent mass $M_c$. After the turnaround the
remanent black hole mass continues to decrease. For black holes with
mass much more massive than the Planck mass, the black hole masses
approach $M_c/2$ asymptotically. Of course, our evaluation after the
turnaround is not very rigid, because, as the contraction of the
universe, the phantom energy will become subdominant in the universe
eventually.

So the remanent mass $M_c$ implies that black holes in the cyclic
model cannot be pulled apart by the phantom energy before
turnaround. Here, we note, if the Friedmann equation
(\ref{Friedmann}) is used, no turnaround occurs, but the big rip.
And then we get
\[
M=\frac{M_i}{1+2DM_i(\sqrt{\rho}-\sqrt{\rho_i})}.
\]
So, in this case, as $\rho\rightarrow\infty$, black holes disappear
$M\rightarrow0$. But, in the cyclic cosmology, the modified
Friedmann equation (\ref{ModFriedmann}) is used. The big rip is
avoid and then, through the phantom accretion, black holes can not
be eliminated. This result can be obtained in another way. Only
using Eq.(\ref{dMdt}), we can obtian
\[
\frac{1}{M}-\frac{1}{M_i}=-\int^{t}_{t_i}{4\pi A(\rho+p)dt}
\]
The black hole mass $M$ will not be zero unless the integration on
the right-hand side is divergent. However, A key property of the
cyclic cosmology is that the energy density and pressure are always
well defined. So it is impossible for black holes in the cyclic
model to be eliminated by accreting the phantom energy.

However, fortunately, we find the remanent masses of black holes at
turnaround do not cause problems. The reason is that these remanent
black holes Hawking evaporate in a time $\tau\sim10^{-43}$.

So our analysis indicates that, although, through the phantom energy
accretion, black holes do not disappear before turnaround, they do
not cause problems in the cyclic models with phantom turnarounds.

\end{document}